# Origins Vs. fingerprints of the Jahn-Teller effect in d-electron ABX$_3$ perovskites

J. Varignon[1,2], M. Bibes[1] and Alex Zunger[3]

[1]Unité Mixte de Physique, CNRS, Thales, Université Paris Sud, Université Paris-Saclay, 91767, France

[2]Laboratoire CRISMAT, CNRS UMR 6508, ENSICAEN, Normandie Université, 6 Bd Maréchal Juin, F-14050 Caen Cedex 4, France

[3]Energy Institute, University of Colorado Boulder Colorado 80309, Boulder, CO, USA

The Jahn-Teller (JT) distortion that can remove electronic degeneracies in partially occupied states and results in systematic atomic displacements is a common underlying feature to many of the intriguing phenomena observed in 3d perovskites, encompassing magnetism, superconductivity, orbital ordering and colossal magnetoresistance. Although the seminal Jahn and Teller theorem has been postulated almost a century ago, the origins of this effect in perovskite materials are still debated, including propositions such as super exchange, spin-phonon coupling, sterically induced lattice distortions, and strong dynamical correlation effects. Although the *end-result* of JT distortions often include a mix of such various contributions, due to coupling of various lattice, spin, and electronic modes with the distortions ("fingerprints", or "consequences" of JT), it is not clear what is *the primary cause*: Which cases are caused by a pure electronic instability associated with degeneracy removal, as implied in the Jahn-Teller theorem, and which cases originate from other causes, such as semiclassical size effects. Here, we inquire about the *origin and predictability* of different types of octahedral deformation by using a Landau-esque approach, where the orbital occupation pattern of a symmetric structure is perturbed, finding if it is prone to total-energy lowering electronic instability or not. This is done for a systematic series of ABX$_3$ perovskite compounds having 3d orbital degeneracies, using the density functional approach. We identify (i) systems prone to an electronic-instability (a true JT effect), such as KCrF$_3$, KCuF$_3$, LaVO$_3$, KFeF$_3$ and KCoF$_3$, where the instability is independent of magnetic order, and forces a specific orbital-arrangement that is accommodated by a BX$_6$ octahedral deformation with a specific symmetry. On the other hand, (ii) compounds such as LaTiO$_3$ and LaMnO$_3$ with delocalized d states do not show any electronically driven instability. Here, their octahedral deformation mode results from coupling of lattice mode with semiclassical size-effects (sterically induced), such as BX$_6$ octahedra rotations. (iii) Although RVO$_3$ (R=Lu-La, Y) perovskites exhibit similar hybridizations as LaTiO$_3$, their t$_{2g}^2$ electronic structure is highly unstable and preserves the Jahn-Teller effect. However, here coexisting steric deformations and JT distortions result in strongly entangled spin-orbital properties. This work provides a unified and quantitative DFT explanation of the experimentally observed trends of octahedral deformations in ABX$_3$ perovskites, without recourse to the dynamically correlated vision of electron interactions codified by the Mott-Hubbard mechanism.

`

## I. Introduction

ABX$_3$ (X=O, F) perovskites[1,2] show a number of systematic atomic distortions relative to the ideal cubic perovskite structure: the one made of corner-sharing, vertically positioned, all parallel BX$_6$ octahedra with equal B-X bonds. The interpretation of much of the electronic and magnetic phenomenology surrounding such perovskites[1,2], including superconductivity, colossal magnetoresistance, orbital ordering and metal-insulator transitions is intimately related to the understanding of the *causes vs. consequences* of the observed atomic distortions. Usually, distortion in ABX$_3$ perovskites produce inequivalent B-X bonds length, such as the Q$_2$ motion (*cf.* Figure 2 of Ref.[3]) that differentiate bond length along the x and y directions. These exist either as in-phase motions in consecutive planes along the z-axis (here labelled Q$_2^+$ motion, Fig1.c) or as in anti-phase along -z (here labelled Q$_2^-$ motion, Fig1.d). Although octahedral deformations tilts (e.g., the in-phase $\emptyset_z^+$ and anti-phase $\emptyset_z^-$, displayed in Fig1.a, b) in different ABX$_3$ compounds are often similarly looking, they might have different origins. Possibilities include being (i) a consequence of the classic *size mismatch* between the A, B, and X atoms, as reflected by geometrical packing constructs such as the Goldschmidt tolerance factor[4], or (ii) a true Jahn-Teller (JT)[1,3,5,6] effect, reflecting an *electronic instability* associated with partial occupation of degenerate states, as described in the seminal work of Jahn and Teller[6]. Understanding predictively the causes of the observed displacements is central to interpreting numerous related phenomena *e.g.* gap formation due to symmetry lowering, metal-insulator transitions[7], "orbital-orderings"[8–10] and magnetic interactions[11] in ABX$_3$ perovskites. This is important since controlling octahedra deformations by using temperature, ferroelectricity, strain engineering or heterostructures[12–19], may offer much needed knobs to control electronic devices.

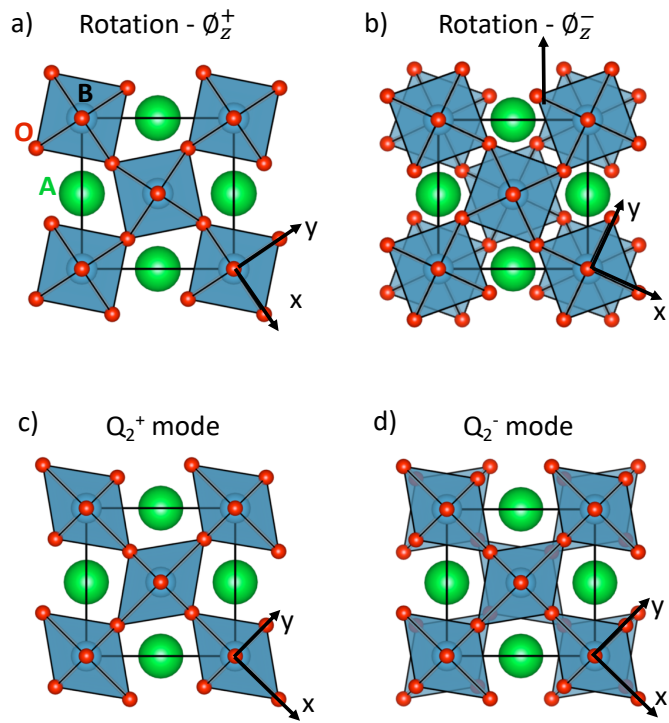

**Figure 1: Sketches of the key lattice distortions appearing in ABX$_3$ materials.** a and b) In phase ($\phi_z^+$, a) and anti-phase ($\phi_z^-$, b) octahedral rotations around the z axis. c and d) Octahedral deformations propagating in phase ($Q_2^+$, c) or in anti-phase ($Q_2^-$, d) along the z axis.

***Puzzles:*** Predictive understanding of the mechanisms of octahedral distortions is rather challenging, for a number of reasons:

(a) Compounds with the same degenerate electronic configurations such as KCrF$_3$[20], KCuF$_3$[21] and LaMnO$_3$[22] being all $e_g^1$; or YVO$_3$ and LaVO$_3$ being both $t_{2g}^2$ (Ref. [23]) show different types of octahedral deformations at low temperature: $Q_2^-$ for KCrF$_3$, KCuF$_3$ or LaVO$_3$, and $Q_2^+$ for YVO$_3$ and LaMnO$_3$, raising the question of what creates $Q_2^+$ and what creates $Q_2^-$;

(b) Recent theoretical works predicted the surprising appearance of strong in-phase $Q_2^+$ symmetry-lowering octahedral deformations in compounds such as SrTiO$_3$, BaMnO$_3$ or BiFeO$_3$ even when there are no degenerate states ($t_{2g}^0$, $t_{2g}^3$ or $t_{2g}^3e_g^2$ configurations, respectively)[14,24];

(c) The textbook depiction of the JT force is often based on calculations based on the degenerate *orbital alone,* but one must consider the force resulting from the systems total energy, including not only the sum of one-electron orbital energy but also the electron-electron Coulomb and exchange-correlation energies resulting from degeneracy removal.

*Literature models* for explaining the origin of octahedral deformations in ABX$_3$ materials span a large range of mechanisms, including electronic super-exchange as codified by the Kugel-Khomskii model [10,11], electron-phonon coupling[9,25], lattice mode couplings involving rotations and forcing the $Q_2^+$ mode [13,26,27], or specific dynamical correlation effects [28,29]. The Kugel-Khomskii approach represents a phenomenological approach to understand the connection between orbital and spin order. However, it does not identify electronic and structural origins of what is a Jahn-Teller effect in perovskite, since both $Q_2^+$ and $Q_2^-$ modes are equally assumed to be JT distortions. Regarding the crucial role of dynamical correlations on the stabilization of JT distortions in perovskites, previous work claimed that *LDA and GGA* "*usually fail to predict the correct electronic and structural properties of materials where electronic correlations play a role*" [9], similar statements also abound in literature[9,28,29]. Nevertheless, such a belief was often grounded on a naïve implementation of DFT using either non-spin polarized simulations, and/or lack of symmetry breaking modes and use of simple exchange-correlation functionals that do not distinguish occupied from unoccupied orbitals. Recent works [30–33] have clarified the fact that strong dynamical correlations are not the determining factor neither for gap opening not for symmetry lowering displacements in ABO$_3$ materials.

Since none of these existing models or visions to explain JT distortions have been applied to a full range of ABX$_3$ compounds with different $e_g$-like and $t_{2g}$-like orbital occupations, at this time it is not clear if different mechanisms are needed to explain different distortions in different compounds, if $Q_2^+$ and $Q_2^-$ modes really reflect Jahn-Teller effects and if there is a basic theoretical framework that could explain them all.

***Distinguishing features of the current approach:*** We address this subject by studying several ABX$_3$ materials (X=O, F) with a transition metal element exhibiting degenerate states (but without propensity to undergo disproportionation effect) using mean field like Bloch periodic DFT band theory with two stipulations. First, a polymorphous representation of the real space structure that permits the existence of different local environments to atoms and spins is required. This entails using crystallographic cells that are not limited to the smallest primitive cell, and therefore allow orbital, spatial and spin symmetry breaking, should these modes lower the total energy. Second, breaking orbital symmetries of degenerate partners requires an exchange correlation functional that distinguishes occupied from unoccupied states, thereby affording significant cancellation of the self-interaction error and thus spatial compactness of 3d orbitals. Here we use DFT+U, (but nonzero U is not essential, as other exchange correlation functional make no use of an explicit U term[30]). This approach provides DFT with a fair opportunity to reveal if the different patterns of symmetry breaking—spin order; JTD and /or ORT, and orbital occupation symmetry breaking—lower the energy or not, thereby establishing a first-principles framework for predictive theory of displacements in ABX$_3$ perovskites.

*A Landau-esque perturbation approach:* To determine if a potential deformation has electronic origin or not we use a Landau-esque perturbation approach that examines whether symmetry lowering via nudged occupation numbers in a degenerate manifold of an initially high symmetry cubic cell lowers the total energy or not. We examine the total energy for a number of prototypical cases encountered in 3d ABX$_3$ compounds: (a) equal occupations scenario of a degenerate level [such as (1/2, 1/2) for a single electron in doubly occupied e$_g$ level] and (b) Orbital Broken Symmetry (OBS) scenario of degenerate partners, *e.g.* (1, 0). If (b) gives total energy lowering relative to (a) this signals the propensity for an electronic instability in the parent high symmetry phase, which is then followed up by the complete relaxation of the unit cell displacements, providing our predicted JT distortions (possibly with coupling to octahedral tilting). If the unequal occupations configuration (b) returns to the equal occupation configuration (a) during self-consistent calculation, this indicates that the high symmetry system is electronically stable and thus not prone to develop an *electronically enforced* JT distortion. Distortions present may reflect other factors such as the classic steric effects induced by size mismatch.

*The main conclusions:*

(a) The existence of electronic instabilities, *i.e.* a true JT effect is predicted in the high symmetry Pm-3m cubic phase of several ABX$_3$ compounds (KCrF$_3$, KCuF$_3$, LaVO$_3$, KFeF$_3$ and KCoF$_3$) whereas LaMnO$_3$ or LaTiO$_3$ have no such instability. Instabilities are manifested by breaking of orbital degeneracies while lowering the total energy, and thereby concomitantly opening finite band gaps.

(b) "Orbital-ordering" is always found to be a total energy lowering event. In such a state, electrons occupy an orbital that is pointing to orthogonal directions between all nearest-neighbor 3d atoms, such as alternation of d$_{xy}$/d$_{yz}$ orbitals for a material with a t$_{2g}$ degeneracy or alternation of $d_{y^2}/d_{x^2}$ orbitals for a material with an $e_g$ level degeneracy, thereby minimizing orbital interactions as codified by the phenomenological electronic super-exchange model of Kugel and Khomskii. The above observations are independent of the imposed magnetic ordering, thus substantiating the view the magnetic order is not the cause of the Jahn-Teller effect and related orbital-ordering, but rather a consequence of these phenomena.

(c) The $Q_2^-$ octahedral deformation mode is a Jahn-Teller distortion: Following the identification of electronic instabilities in cubic cells, we follow the quantum mechanical forces to establish the fully relaxed crystal structures, finding that the aforementioned JT distorted compounds develop a $Q_2^-$ octahedra deformation mode, which is therefore identified as an electronically induced Jahn-Teller distortion. The significance of the development of this specific $Q_2^-$ mode is that it is characterized by *opposite* octahedral deformations between nearest sites and is thus able to accommodate the electronically induced orbital-ordering. Calculation of the magnitude of the displacements shows good agreement with the experimentally observed trends in a full range of $ABX_3$ compounds with $t_{2g}$ and $e_g$ degenerate partners

(d) At odds with the common perception, $LaMnO_3$ does not exhibit JT effect whereas $KCrF_3$ does, despite isovalent ($t_{2g}^3 e_g^1$) configurations. Similarly, the $t_{2g}^1$ configuration $LaTiO_3$ has no JT distortion, whereas $KFeF_3$ does. This is because the existence of strong hybridization in the oxide cases diminishes orbital localization needed to create JT distortion, whereas in the fluoride the hybridization is much weaker.

(e) The $Q_2^+$ octahedral deformation, appearing in $LaMnO_3$ or $LaTiO_3$, is not induced by an electronic instability and thus it is not a Jahn-Teller distortion. It is a consequence of octahedral rotations and tilts often appearing in the perovskite oxides due to pure semiclassical atomic size effects, i.e. geometric steric effects. The $Q_2^+$ octahedral deformation mode thus cannot signal strong dynamical correlation effects in these $ABX_3$ compounds.

(f) Although $LaVO_3$ exhibits similar B, d – X, p hybridizations as $LaTiO_3$, the former compound exhibits a robust electronic instability in the cubic cell while $LaTiO_3$ has zero stabilization energy. The reason is that $LaVO_3$ has two electrons $t_{2g}^2$ relative to $LaTiO_3$ with just one $t_{2g}^1$.

(g) Due to the coexistence of a JT effect and of octahedra rotations in $RVO_3$ materials, $Q_2^-$ and $Q_2^+$ modes compete and result in two distinct spin-orbital orders at low temperature depending on the octahedra rotations amplitude.

(h) JT distortions as well as semiclassical size effect distortions can contribute to the opening of band gaps in Mott insulators. We have previously identified[33] four gapping modalities; the JT effect contributes just to modalities (iii) as follows: (i) compounds with closed subshells can open a gap due to octahedral crystal field splitting ($CaMnO_3$, $LaFeO_3$), (ii) compounds opening gaps by lifting degeneracies through large symmetry lowering displacements such as $X_6$ rotations ($LaTiO_3$ or $LaMnO_3$), (iii) compounds with two electrons in $t_{2g}$ levels exhibiting Jahn-teller induced electronic instability able to cause gapping such as $LaVO_3$ ($t_{2g}^2 e_g^0$) and (iv) compounds with unstable single local electronic occupation patterns disproportionating into a double local environment *e.g.* $CaFeO_3$ and $YNiO_3$.

We conclude that the electronically-induced Jahn-Teller distortion mode $Q_2^-$ and the geometrically induced steric $Q_2^+$ octahedral deformation mode are fully captured by a static mean-field method. This is in line with recent theoretical works that have demonstrated that static mean-field methods capable of inducing broken symmetry such as Density Functional Theory[13,26,30,33] in a polymorphous representation suffice to also explain (i) the trends in gapping and type of magnetic order across the ternary $ABO_3$ series[33] and the binary 3d oxide series[31,34]; (ii) the trends in disproportionation into two different local environments of the B site $2ABO_3 \rightarrow A_2[B,B']O_6$[32] and (iii) the explanation of doping Mott insulators including cuprates[35,36], doping Kagome structures[37] as well as "anti-doping" oxides[38].

## II. The elements of the method

The main features of the theoretical framework used in the study are as follows:

**(a)** *Cell geometry and relaxation:* We consider the following structure types: high symmetry Pm-3m cubic cell ($LaTiO_3$, $LaVO_3$, $LaMnO_3$, $KFeF_3$, $KCoF_3$, $KCrF_3$ and $KCuF_3$) as well as the experimentally observed structures, namely I4/mcm ($KCuF_3$, $KCrF_3$), Pbnm ($LaTiO_3$, $LaVO_3$, $LaMnO_3$, $LaTiO_3$), $P2_1/b$ ($LaVO_3$), $I2/m$ ($KCrF_3$), P-1 ($KCoF_3$) and $I2/a$ ($KFeF_3$) for identifying the DFT ground state structure. We allow cell sizes larger than the minimal (one formula unit per cell) so as to permit symmetry breaking distortions, should they lower the total energy. Atoms have to be nudged initially off their high symmetry positions in the cubic cell, followed by the calculation of the restoring Hellman Feynman forces guiding full relaxation. In order to provide sufficient flexibility for developing general patterns of energy lowering deformations we have used a polymorphous crystallographic cell corresponding to a (√2a, √2a, 2a) cubic cell (*i.e.* 4 f.u per cell). The structural relaxation (lattice parameters and atomic positions) of ground state structures has also been performed until forces are lower than 1 meV/Å. We performed a symmetry adapted mode analysis that allows to extract the amplitudes of general distortion modes by projecting the distorted structure on the basis of the phonon eigen-displacements of the high symmetry cubic cell[39,40].

**(b) *Orbital symmetry breaking:*** The occupation numbers of partially filled degenerate orbitals in the high symmetry cubic cell is not restricted to equal occupations of the degenerate partners. For a compound with a single $e_g$ electron, we do not pre-select a (1/2, 1/2) occupation pattern as generally done in standard band calculations, but allow also exploration of a (1,0) occupation pattern. Analogously, for 2 electrons in the $t_{2g}$ level we explore (2/3, 2/3, 2/3) *vs.* (1,1,0) occupation patterns. Wave function symmetrization to a presumed symmetry is avoided so as to allow electrons to freely occupy energy lowering configurations. We therefore initially nudge not only atomic positions but also the occupation patterns of electrons in specific degenerate partners, called orbital broken symmetry states (OBS) [41–43]. The self-consistent field with possible changes in d orbital occupancies is then obtained starting from this initial guess.

**(c) *Spin order:*** We use the observed low temperature spin order, *e. g.* AFMA (ferromagnetic planes coupled antiferromagnetically together), AFMC (AFM planes coupled ferromagnetically together) and AFMG (all nearest neighbor cations are antiferromagnetically coupled). Since we have previously shown that low temperature spin-ordered phase inherits the physics of the high temperature PM phase in $ABO_3$ materials [33], we did not attempt to model the PM distortions presently. For the study of the linear response of the ideal cubic phase, we prefer to use a simple FM order so as to avoid potentially strongly entangled spin-orbital situations as codified by the Kugel-Khomskii model [10,11], *i.e.* the chicken and the egg dilemma. This is reinforced by experimental observations of octahedra deformations appearing at higher temperatures than the AFM to PM (for example, in $KCrF_3$, $KCuF_3$, $LaMnO_3$).

**(d) *Exchange-correlation functional:*** To enable energy lowering occupations one needs to use an exchange-correlation (XC) functional that distinguishes occupied from unoccupied states. Local (LDA) and semi-local (GGA, meta-GGA) XC functionals do not make such distinction. The simplest XC allowing this is DFT+U, where U is an on-site potential acting on a subset of orbitals – here the transition metal (TM) d states – and shifting to lower (upper) energies occupied (unoccupied) levels. We have thus employed this formalism in combination to the PBEsol[44] XC functional where U is an effective parameter $U_{eff}=U-J$.[45] We did not optimize the effective U values for each material to achieve an optimal fit (this might be done if needed in the future), but opt instead to use fixed U=3.5 eV for all 3d TM elements for simplicity.

### III. Results

#### A. Examining the propensity of the symmetric structure for electronically-induced distortions

The Landau-esque symmetry breaking test perturbs a high symmetry cubic cell (here, Pm-3m with lattice parameter $a_{cub}$) and examines the total energy for (a) assumed equal occupations of the degenerate partners [no-OBS, such as (1/2, 1/2) for a single electron in doubly degenerate $e_g$ level] and (b) Orbital Broken Symmetry (OBS), looking for energy lowering configurations [*e.g.* (1, 0)]. The spin configuration used is FM. Energy differences between these configurations are provided in Table I.

We see that given the opportunity for orbital broken symmetry, the compounds LaVO$_3$, KFeF$_3$, KCoF$_3$, KCrF$_3$ and KCuF$_3$ show an energy gain $\Delta E_{OBS-no\ /OBS}< 0$, with a concomitant opening of the bang gap, while LaTiO$_3$ and LaMnO$_3$ have $\Delta E_{OBS-no\ /OBS}= 0$, *i.e.* the initially imposed OBS relaxes back to configuration with equally occupied degenerate orbital, and the system stays metallic. These observations are unchanged by using another non-local functional such as HSE06[46,47] hybrid functional method, having also a good adherence to self-interaction cancelation. It is significant that the sign of the energy difference is unaltered when one uses instead of the FM spin configuration the ground state magnetic order, namely AFMG for LaTiO$_3$, KFeF$_3$ and KCoF$_3$, AFMC for LaVO$_3$ and AFMA for LaMnO$_3$, KCrF$_3$ and KCuF$_3$, respectively, see numbers in parentheses in Table I.

|  | Electronic configuration | ($\sqrt{2}a_{cub}$, $\sqrt{2} a_{cub}$, $2 a_{cub}$) cell (DFT+U) | | Primitive cubic cell (HSE06) |
|---|---|---|---|---|
|  |  | $\Delta E_{OBS- no /OBS}$ (meV/f. u) | $E_g$ (eV) | $\Delta E_{OBS- no /OBS}$ (meV/f.u) |
| LaTiO$_3$ | d$^1$ (t$_{2g\uparrow}^1$) | 0 (0) | 0 | 0 |
| LaMnO$_3$ | d$^4$ (t$_{2g\uparrow}^3$e$_{g\uparrow}^1$) | 0 (0) | 0 | 0 |
| LaVO$_3$ | d$^2$ (t$_{2g\uparrow}^2$) | -297 (-237) | 0.42 | -428 |
| KFeF$_3$ | d$^6$ (t$_{2g\uparrow}^4$e$_{g\uparrow}^2$t$_{2g\downarrow}^1$) | -655 (-873) | 1.78 | -780 |
| KCoF$_3$ | d$^7$ (t$_{2g\uparrow}^4$e$_{g\uparrow}^2$t$_{2g\downarrow}^2$) | -726 (-895) | 1.90 | -1071 |
| KCrF$_3$ | d$^4$ (t$_{2g\uparrow}^3$e$_{g\uparrow}^1$) | -124 (-85) | 0.66 | -256 |
| KCuF$_3$ | d$^9$ (t$_{2g\uparrow}^4$e$_{g\uparrow}^2$t$_{2g\downarrow}^3$e$_{g\downarrow}^1$) | -71 (-81) | 0.38 | -477 |

*Table I: Detection of spontaneous electronic instability in the high symmetry cubic phase of ABX$_3$ perovskites. Energy differences between solutions with equal occupancy of degenerate levels (no OBS) and with the most stable orbital broken symmetries (OBS) state in meV per formula units obtained in GGA+U and HSE06 functionals (the latter uses a smaller cubic cell for computational cost reason). The lattice parameter a is fixed to a relaxed cubic cell without OBS. A FM order is assumed. Results for the ground state AFM order is reported in parenthesis, namely AFMG for LaTiO$_3$, KFeF$_3$ and KCoF$_3$, AFMC for LaVO$_3$ and AFMA for LaMnO$_3$, KCrF$_3$ and KCuF$_3$, respectively.*

1. **The emergence of orbital order as a DFT total energy lowering Jahn- Teller effect**

Once the degeneracies of the high symmetry phase are removed, one notes the creation of deterministic localization patterns of the degenerate partner components (such as d$_{x^2-y^2}$ and d$_{z^2}$) on different atomic sites. This "orbital ordering" has intrigued the community of 3d oxides[3,11], raising various exotic interpretations[10,11,13,25–29]. To understand the possible unexotic, mean field total energy origins of such orbital ordering, we have investigated the energetics of different assumed orbital-arrangement: (a) electrons in the degenerate levels are initially nudged in *identical* orbitals on all nearest transition metal sites, such as a d$_{z^2}$ occupancy on all sites for a compound with a single e$_g$ electron; (b) the electron is initially nudged in a *different* degenerate partner between nearest TM sites in the (xy) plane but on *the same* partner for TM located in the consecutive plane along the z direction, such as alternation of d$_{x^2}$ and d$_{y^2}$ in the (xy) plane with similar arrangement on the consecutive plane along z, thus forming a "columnar arrangement" and (c) the electron is initially nudged in a *different* degenerate partner between *all* nearest neighbor site, such as alternation of $d_{x^2}$ and $d_{y^2}$ in all cartesian directions, thus forming a 3D checkerboard. For a compound with 2 unpaired electrons for $t_{2g}$ levels, case (b) and (c) necessarily require that an identical orbital has to be occupied between nearest TM sites. We thus imposed one electron in a specific $t_{2g}$ partner on all TM sites ($d_{xy}$ here) and then we alternate the occupancy of the remaining partners ($d_{xz}/d_{yz}$). After initial nudging, (intended to avoid accidental local minimum), the solution is iterated to self-consistency with full relaxation of the electronic structure but without any structural relaxation.

The results show that for all compounds displaying a spontaneous electronic instability, *the lowest energy state is always associated with configuration (c) – e.g. alternation of $d_x^2$ and $d_y^2$ in all cartesian directions, thus forming a 3D checkerboard*. (The total energies of single, columnar and checkerboard OBS states are provided in Supplementary section I" Total energies of Orbital Broken Symmetry states in cubic cells"). In such a minimum energy state, electrons occupy an orbital that is pointing to orthogonal directions between all nearest-neighbor transition elements, such as alternation of $d_{xy}/d_{yz}$ orbitals for a material with a $t_{2g}$ degeneracy, or alternation of $d_{y^2}/d_{x^2}$ orbitals for a material with an $e_g$ level degeneracy as shown by our partial charge density maps of states near the Fermi level (Fig 2.a, b, d and e). This specific pattern is referred to as a G-type antiferro-orbital ordering[13,23,48] that minimizes orbital interactions between all nearest neighbor TM site on a perfectly cubic lattice. Significantly, these energy-lowering states all exhibit band gaps with respect to the cubic cell with equally occupied degenerate partners, even though no structural relaxation has been performed as yet (see Table I). *We conclude here that in $LaVO_3$, $KFeF_3$, $KCoF_3$, $KCrF_3$ and $KCuF_3$, the electronic structure in the cubic cell is distorting in order to remove the orbital degeneracy, thereby producing an antiferro-orbital arrangement and opening a band gap. These compounds therefore exhibit the signatures of a Jahn-Teller effect that is directly related to gapping.*

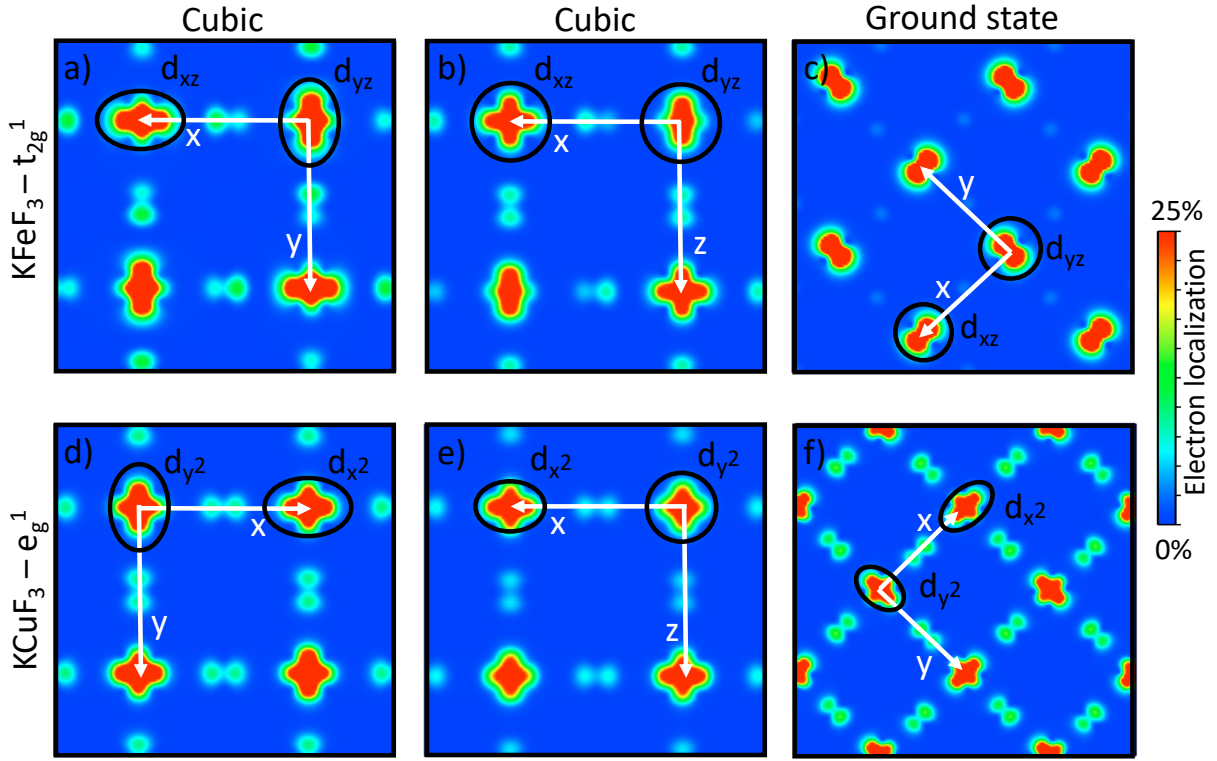

*Figure 2: Orbital orderings appearing ABX₃ materials. Wavefunction squared maps of electrons located at the top of the valence bands (latest two occupied bands) for KFeF$_3$ (a, b and c) and KCuF$_3$ (d, e and f) in their unrelaxed cubic cells (a, b, d and e) and in the fully relaxed ground state structure (c and f).*

**2. When electronic delocalization of degenerate orbitals prevents the JT effect**

One might wonder at this stage why compounds with identical electronic degeneracy differ in their ability to have a JT effect. For example, LaMnO$_3$ does not exhibit a JT effect whereas KCrF$_3$ does; similarly, LaTiO$_3$ has no JT effect, whereas KFeF$_3$ does. To understand this, let us recall that the energy surface of a JT system[49] as a function of a displacement x can be represented as $E(x) = -F_{JT}x + 1/2Kx^2$, consisting of a stabilizing electronic JT force $F_{JT}$ associated with electronic degeneracy removal, and an opposing harmonic restoring force of the surrounding bond characterized by the force constant K. That the existence of degeneracy of partially-filled states does not automatically force a JT distortion is clear from the competition between these two terms: if the relevant (degenerate) orbitals are too hybridized/delocalized (weak $F_{JT}$), or the harmonic response of the bonds about to be deformed is too stiff, degeneracy will not lead to a JT distortion. Strongly (weakly) p-d hybridized systems will tend to have a weaker(stronger) JT force. To assess such tendencies, one must carry out a JT distortion inspecting the total (electron-electron; electron-ion and ion-ion) energy terms, not just orbital energies. This is readily done in DFT supercell theory.

To assess the difference in hybridization between fluorides and oxides, Fig. 3 shows the projected density of states of isoelectronic pairs (LaTiO$_3$ *vs.* KFeF$_3$, as well as LaMnO$_3$ *vs.* KCrF$_3$) in the high symmetry cubic cell with equal occupancies of degenerate partners. As one can see, the fluorine-based compounds show minimal hybridizations between the B-d and X-p states, and the bandwidth associated with degenerate partners is rather narrow, thus resulting in a localized electronic structure and d-d like band edges. On the other hand, for oxygen-based perovskites, the hybridization between B-d and X-p states and the band width associated with degenerate partners increases continuously upon adding electrons to the d levels, until reaching a 'charge transfer insulator' regime for LaMnO$_3$ – *i.e.* anion p like VBM and cation d like CBM. This thus suggests that oxides have more delocalized d states than fluorides. Thus, in the oxides the JT force can be too weak and might be overcome by the restoring force but in the fluorides the JT force is stronger and might win. This is what we see in Table I.

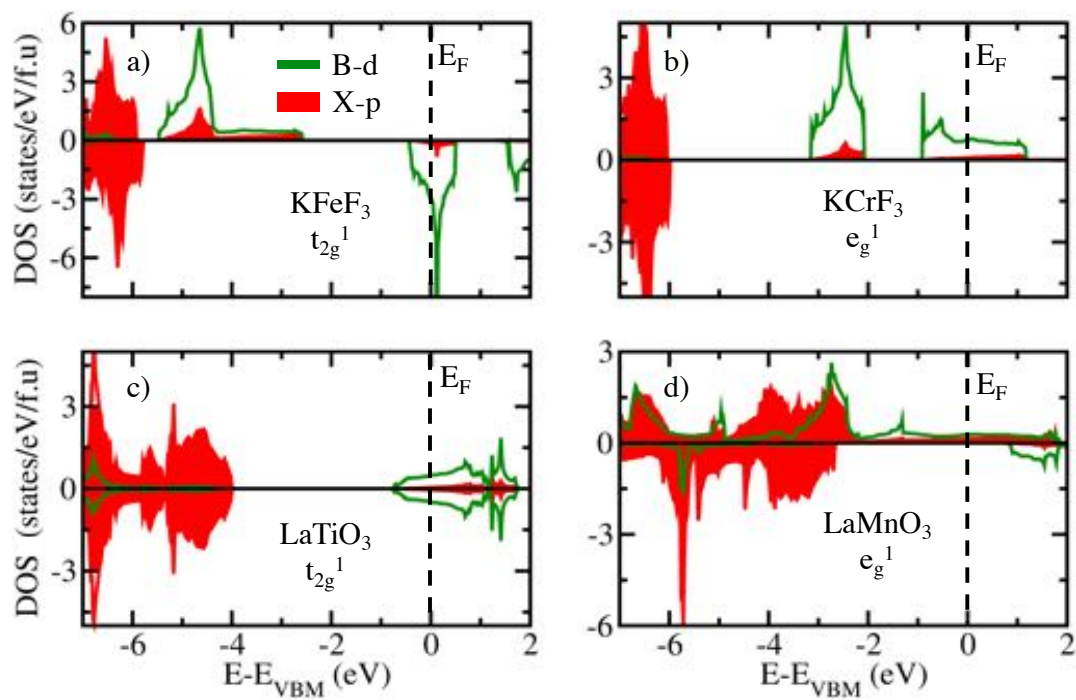

*Figure 3: Electronic structure of compounds with isoelectronic degeneracies. Projected density of states on B-d (green) and X-(red) p states in materials showing a $t_{2g}$ (KFeF$_3$ and LaTiO$_3$, panels a and c) and $e_g$ (KCrF$_3$ and LaMnO$_3$, panels b and d) isoelectronic degeneracies in their cubic cell. The vertical line indicates the Fermi level. A simple FM order is used.*

To next check how the degree of localization of degenerate orbitals induces JT displacements, we design enhanced localization in LaTiO$_3$ by increasing the lattice constant, *i.e.* using a (negative) hydrostatic pressure. Starting from this new cubic cell in which the bandwidth of the t$_{2g}$ degenerate partners is reduced by 0.5 eV *wrt* the unperturbed cubic cell, we again nudge the Ti$^{3+}$ electron to specific t$_{2g}$ partners (see Supplementary Section II "Density of states of LaTiO$_3$ under negative pressure"). LaTiO$_3$ now develops a small electronic instability (ΔE$_{OBS-no\ OBS}$=-6 meV/f.u) producing an antiferro-orbital occupancy between all nearest neighbor Ti$^{3+}$ sites and opening a band gap of 0.1 eV. Therefore, upon decreasing B, d – X, p hybridizations and the t$_{2g}$ levels bandwidth, LaTiO$_3$ undergoes a Jahn-Teller effect.

*We conclude that delocalized degenerate states can prevent an electronic instability and the ensuing Jahn-Teller effect despite the presence of orbital degenerate states.* This also explains why SrVO$_3$ or CaVO$_3$, showing strongly hybridized electronic structures, do not develop a JT effect and thereby stay metallic.

### 3. Trends in JT energies with orbital occupancy

Although LaVO$_3$ exhibits similar B, d – X, p hybridizations as LaTiO$_3$, the former compound possesses a robust electronic instability in the cubic cell (energy lowering of 297 meV; Table I), while LaTiO$_3$ has zero stabilization energy. This reflects the fact that the JT force is larger in systems with higher orbital occupancy – LaVO$_3$ with two electrons t$_{2g\uparrow}^2$ relative to LaTiO$_3$ with just one t$_{2g\uparrow}^1$. Similarly, the energy gain associated with breaking orbital degeneracies increases when going from t$_{2g}^1$ in KFeF$_3$ to a t$_{2g}^2$ in KCoF$_3$. In general, we expect that triply degenerate states occupied by 2 electrons to have the greatest JT instability.

We next turn to study the trends in the deformation amplitudes.

## B. The Q$_2^-$ octahedral deformation mode is a Jahn-Teller distortion

### 1. Electronically unstable cubic structures develop the Q$_2^-$ octahedral deformation upon relaxation

While literature often ascribe Q$_2^+$ and Q$_2^-$ modes to a JT effect [2,3], we inspect here the type of distortion forced by the spontaneous electronic instability in the hypothetical cubic structure of $LaVO_3$, $KFeF_3$, $KCoF_3$, $KCrF_3$ and $KCuF_3$. To that end, we allow these systems to change their structures by fully developing energy lowering displacements. Application of structural relaxation techniques (following quantum mechanical forces to zero force configurations) reveals all these systems develop specifically the anti-*phase $Q_2^-$ octahedral deformatio*n that we therefore consider as the fingerprint of JT effect in these systems. We tested the $Q_2^+$ octahedral deformation mode but this octahedral deformation mode produces lower energy gains than the $Q_2^-$ mode (see Supplementary Section III " Energy gain associated with $Q_2^+$ and $Q_2^-$ octahedra deformation mode").

Symmetry adapted mode analysis of the relaxed DFT structures (starting from experimentally observed structures and magnetic orders) are presented in Table II. One can appreciate the development of the specific $Q_2^-$ mode by the presence of shifted single well potential energy surface associated with the $Q_2^-$ mode starting from a cubic cell, signaling the presence of a Jahn-Teller force $F_{JT}$ (see Supplementary Section IV "Potential energy surfaces associated with the $Q_2^-$ "). Note that the shifted single wells are independent of the spin order – FM and AFM orders yield similar results—, and thus the magnetic order at low temperature may be a consequence of the specific orbital-orderings forced by the electronic instability, but not its cause. The significance of the development of this specific $Q_2^-$ mode (Figure 1.d) is that it is characterized by *opposite* octahedral deformations between nearest sites, and is thus able to accommodate and amplifies the incipient orbital-orderings (Fig 2.c and f). Specifically, the B-X bonds expand and contract in the plane defined by the alternating directions of the occupied orbitals.

We emphasize that the appearance of an anti-phase rotation propagating orthogonally to the JT distortion leads to very small additional components to the symmetry allowed JT distortion in which apices O move inward/outward (Supplementary Section V explains "The anti-phase rotation and the $Q_2^-$ motion displacements patterns describe orthogonal planes in $KCrF_3$ for minimizing electronic super-exchange"). Although it is sometimes proposed to come from additional electron interactions[50], it is just a consequence of the $X_6$ rotation (see Supplementary Section VI "Low temperature orbital and structural transition in $KCrF_3$").

2. **Theory vs. experimental observation for the $Q_2^-$ octahedral deformation mode**

The predicted specific symmetry of the $Q_2^-$ mode is precisely what is needed to explain the cubic to tetragonal $I4/mcm$ structural transition in $KCrF_3$ and $KCuF_3$ (observed at T=973 K and 800 K[51,52], respectively), and of the Pbnm to $P2_1/b$ structural transition observed in $RVO_3$ compounds (R=Lu-La, Y)[23,53]. Table II summarizes the calculated *vs.* experimental values of the *amplitudes* of the $Q_2^-$ displacements, showing very good agreement.

| | t factor | Magnetic order | Space Group | $Q_2^+$ ($M_3^+$) Calculated (Exp.) | $Q_2^-$ ($R_3^-$) Calculated (Exp.) |
|---|---|---|---|---|---|
| **LaTiO$_3$** | 0.93 | AFMG | Pbnm | 0.040 (0.041[54]) | - |
| **LaMnO$_3$** | 0.94 | AFMA | Pbnm | 0.324 (0.357[22]) | - |
| **LaVO$_3$** | 0.95 | AFMC | P2$_1$/b | 0.005 (0.009[55]) | 0.093 (0.079[55]) |
| | | | Pbnm | 0.078 (0.090[56]) | - |
| **KFeF$_3$** | 1.00 | AFMG | I$_2$/a | - | 0.104 (-) |
| **KCoF$_3$** | 1.01 | AFMG | P-1 | - | 0.003 (-) |
| **KCrF$_3$** | 0.99 | AFMA | I$_2$/m | - | 0.336 (0.316[20]) |
| | | | I$_4$/mcm | - | 0.300 (0.299[20]) |
| **KCuF$_3$** | 1.03 | AFMA | I$_4$/mcm | - | 0.335 (0.355[21]) |

*Table II: Amplitudes associated with structurally distorted octahedra in the ground state of ABX$_3$ perovskites. Amplitudes of distortions (in Å) associated with the $Q_2^+$ and $Q_2^-$ modes (irreducible representations $M_3^+$ and $R_3^-$ respectively) distorting octahedra of the optimized structures starting from a cubic cell with the A cation located at the corner of the cell. Experimental values extracted from structures available in literature are provided in parenthesis. The Goldschmidt t factor is also reported, as well as the magnetic state observed experimentally at low temperature and assumed in the simulation for the relaxation of the ground states.*

## C. The $Q_2^+$ motion in LaTiO$_3$ and LaMnO$_3$ is a consequence of classic octahedral rotations and tilting, not an electronic JT effect

Consistent with the absence of electronic instabilities in the high symmetry Pm-3m cubic cell of LaTiO$_3$ and LaMnO$_3$, we find that the Q$_2^+$ or Q$_2^-$ octahedra deformation are associated with a single well energy potential (see figures 4.a and b) and are unable to open a band gap. Such observations were already raised by Lee *et al* for LaMnO$_3$ in Ref.[57]. This means that, although a sizable Q$_2^+$ octahedra deformations mode appears in the ground state structure of LaTiO$_3$ and LaMnO$_3$, this mode is not produced by an electronic instability, does not lift orbital degeneracies or produce the metal-to-insulator transition, all being key aspects of the Jahn-Teller effect.

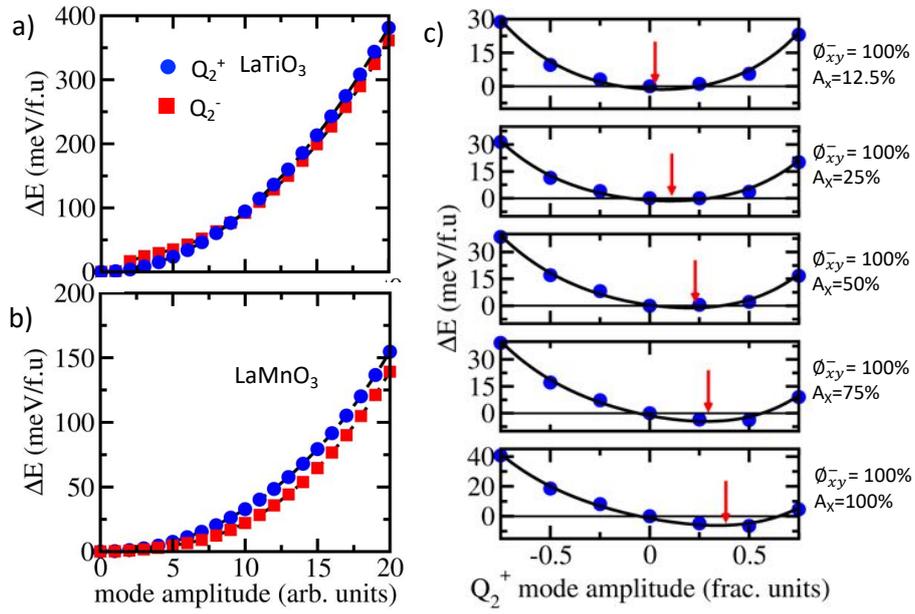

*Figure 4: Induced Q$_2^+$ motions through lattice mode couplings.* a and b) Energy versus Q$_2^+$ (filled blue circles) and Q$_2^-$ (filled red squares) mode amplitudes (arbitrary units) in LaTiO$_3$ (a) and LaMnO$_3$ (b) using a FM order starting from a cubic cell. c) Energy difference (in meV/f.u) versus Q$_2^+$ mode amplitude at fixed amplitude of $\emptyset_{xy}^-$ and A$_X$ in LaMnO$_3$ using a FM order. 1 (100%) represents the amplitude appearing in the ground state. The reference energy is set at 0 amplitude of the Q$_2$ mode.

Nevertheless, these compounds develop two anti-phase and one in-phase X$_6$ rotations (due to low t factors, see Table II) that allow a specific lattice mode coupling between the Q$_2^+$ mode plus the antiphase rotation $\emptyset_{xy}^-$ and the antipolar motion A$_X$ (sketched in Fig 1. of Ref.[26]) of the form $F \propto \emptyset_{xy}^- A_X Q_2^+$ in the free energy expansion as identified in Ref.[13] The appearance of finite amplitudes of $\emptyset_{xy}^-$ and the A$_X$ modes automatically forces finite amplitudes of the Q$_2^+$ mode in order to lower the total energy as we see in Fig.4.c for LaMnO$_3$ .A similar result is found for LaTiO$_3$ (not shown). Although the dependence of the Q$_2^+$ mode on rotation amplitude in LaMnO$_3$ was already shown in Ref.[57], we show that the Q$_2^+$ is nothing but a consequence of ABX$_3$ distortions originating from pure steric atomic size effects captured already by the 1926 pre quantum Goldschmidt tolerance factor[4]. In analogy with improper ferroelectrics[58,59], the Q$_2^+$ mode can be thought of as an improper mode, being a result of a specific octahedra rotations pattern: a combination of two anti-phase and one in-phase rotations. In contrast to often articulated statements[9,22,28,60,61] this Q$_2^+$ motion should not be confused with a JT distortion originating from an electronic instability. The improper mode origin of Q$_2^+$ is consistent with previous theoretical reports of strong Q$_2^+$ mode amplitude reported in compounds that lack degenerate levels[14], clearly signaling that this mode cannot originate from an electronic instability.

1. **The appearance of an Q$_2^+$ mode is not a statement of strong dynamical correlation effects**

Leonov *et al*[28] highlighted the importance of dynamical correlations in stabilization of the Q$_2^+$ mode in LaMnO$_3$. They plotted the potential energy surface as a function of the Q$_2^+$ mode amplitude with Dynamical Mean Field Theory (DMFT) simulations starting from a cell with rotations and the A$_X$ mode, explaining that "*in the calculation we change only the parameter $\partial_{JT}$ (i.e. the amplitude of the Q$_2^+$ mode) … and keep the value of the MnO$_6$ octahedron tilting and rotation fixed*". As can be seen by comparing our mean field DFT result of Fig 4.c with their DMFT results (Fig. 8 of Ref.[28]), the two agree closely, suggesting that in this case, adding dynamic electronic correlations does not bring any new features to the understanding of the octahedra deformation in ABX$_3$ materials.

2. *The origin of the improper Q$_2^+$ motion in LaMnO$_3$*

Our first-principles results demonstrate that the Q$_2^+$ mode is a simple consequence of octahedra rotations, but one wonders if one can extract experimental evidences of this phenomena. At high temperature, LaMnO$_3$ adopts a rhombohedral R-3c phase that is characterized by an anti-phase rotation $\emptyset_{xyz}^-$ around all cartesian axes and in which no octahedral deformation nor related phenomena such as an orbital-ordering are reported [22,62]. This is compatible with our models since (i) no electronic instability yielding a $Q_2^-$ JTD mode is identified in LaMnO$_3$ and (ii) such X$_6$ octahedra tilt pattern does not allow A$_X$ and $Q_2^+$ modes to develop by symmetry. Once LaMnO$_3$ transforms to the Pbnm cell around 750 K, one observes that experimental structures at various temperature taken from Ref.[22] develop an A$_X$ distortion whose amplitude increases with decreasing temperature, *while $\emptyset_{xy}^-$ is not temperature dependent.* Upon increasing of the A$_X$ mode amplitude on cooling, the $Q_2^+$ mode amplitude increases (See supplementary section VII "Symmetry Mode Analysis of LaMnO$_3$ experimental structures" for symmetry adapted modes of these structures). This thus confirms the "improper appearance" of the $Q_2^+$ octahedra distortion mode.

### D. Competing Jahn-Teller effect and sterically induced $Q_2^+$ mode is at the core of the entangled spin-orbital properties in RVO$_3$ compounds

Although $Q_2^+$ and $Q_2^-$ distortions have totally different origins, they can coexist as long as the $Q_2^+$ mode is allowed by symmetry. This is the case in LaVO$_3$ for which one observes finite amplitudes of the two modes in the low temperature phase both at the theoretical and experimental levels (see Table II), although the JT distortion largely dominates. However, among the RVO$_3$ (R=Lu-La, Y) compounds, LaVO$_3$ shows the smallest octahedra rotations. One may question *what happens for a compound showing larger X$_6$ rotations such as YVO$_3$?*

Consistently with previous litterature[13,53], full structural relaxation (atomic position + cell parameters) of YVO$_3$ yields a Pbnm ground state showing only the $Q_2^+$ mode at 0 K. Interestingly, through appearance of the $Q_2^+$ mode, a columnar orbital pattern is stabilized instead of the checkerboard pattern associated with the JT effect. Consequently, super-exchange paths are enabled along the z axis thereby creating AFM interactions along the "columns". Such a behavior is indeed verified in our simulations: when starting from a $(\sqrt{2}a, \sqrt{2}a, 2a)$ cubic cell of YVO$_3$ with specific initial electron nudging (similar to those presented in Section III.A), we find after variational self-consistency that the columnar orbital pattern is more stable than the checkerboard arrangement when considering the G-type AFM order ($\Delta E_{che-col}$= +31 meV/f.u), and conversely for the C-type AFM order ($\Delta E_{che-col}$= -74

meV/f.u). In other words, each $Q_2$ mode is associated to a specific orbital pattern, and consequently to a precise spin order.

Nevertheless, the checkerboard orbital pattern remains the global energy minimum in the cubic cell (between the G and C type AFM orders), a similar conclusion being drawn with a FM order ($\Delta_{Eche-col}$= -34 meV). Therefore, YVO$_3$ should exhibit the signature of a JT effect although the low T phase only shows the $Q_2^+$ mode. This is verified experimentally by the presence of an intermediate P2$_1$/b symmetry characterized by the $Q_2^-$ mode as shown by the symmetry adapted mode analysis presented in Table 1 of Ref.[13]. Thus, the question is now what is the driving force of the transition to the purely orthorhombic *Pbnm* cell at 0 K for YVO$_3$.

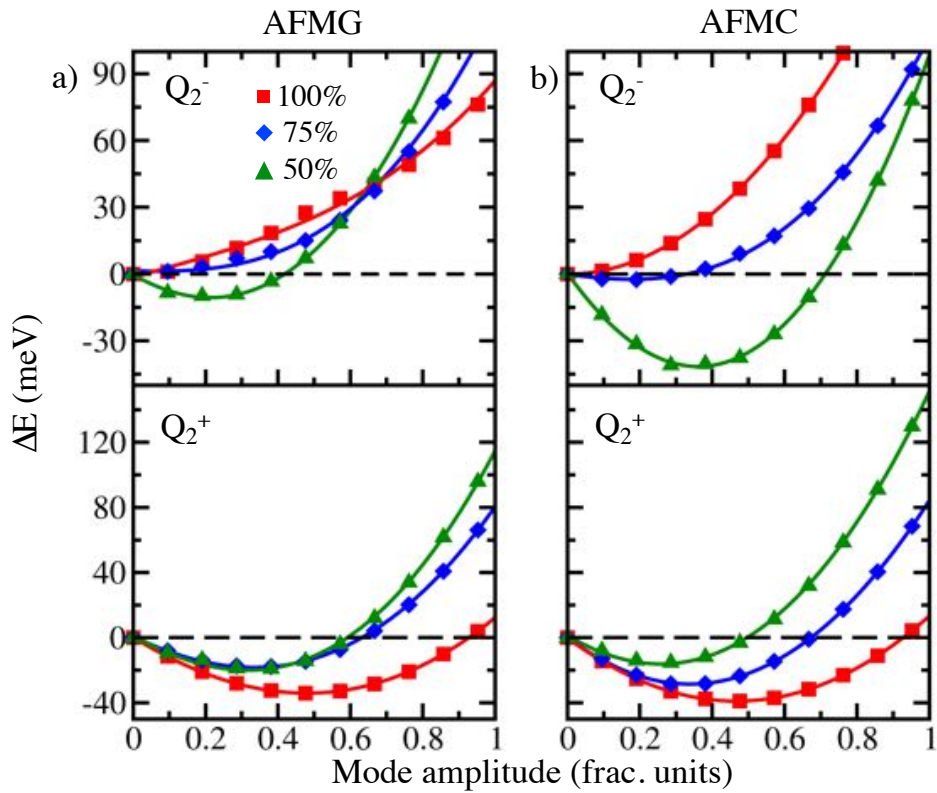

*Figure 5: Energy difference (in meV per 20 atom unit cell) as a function of the $Q_2^+$ and $Q_2^-$ mode amplitudes (in fractional units) at fixed rotation and antipolar displacement amplitude (in %) appearing in the ground state structure of YVO$_3$ using the AFMG (a) and AFMC (b) order. The unit cell is fixed to a ($\sqrt{2}a, \sqrt{2}a, 2a$) cell whose lattice parameter yields the ground state volume. 1 is actually the amplitude appearing in the ground state structure.*

To get insights on this peculiar transition, we start from the high symmetry $(\sqrt{2}a, \sqrt{2}a, 2a)$ cubic cell of YVO$_3$ and we freeze fixed amplitudes of all distortions appearing in the ground state structure except the $Q_2^+$ mode, *i.e.* add only octahedra rotations and antipolar motions of ions. Then, we compute the potential energy surface associated with $Q_2^+$ or $Q_2^-$ modes (see Figure 5). Upon increasing "orthorhombic distortions", two antagonist effects are observed irrespective of the AFMG or AFMC magnetic order: (i) the $Q_2^+$ mode describes a single well potential whose energy minimum is progressively shifted to larger amplitudes while (ii) the $Q_2^-$ mode vanishes. The former observation is in line with the improper origin of the $Q_2^+$ distortion discussed in section C but the latter finding indicates that orthorhombic distortions produce a crystal field sufficient to split the $t_{2g}$ level degeneracies[33] but also diminish electronic super exchange interaction at the core of the JT effect (see Supplementary Section VIII "Cooperating/competing octahedra rotations and Jahn-Teller effect in LaVO$_3$").

Our results are closely compatible with the experimental phase diagram of rare-earth vanadates[23,53] and settle the issue that $Q_2^-$ and $Q_2^+$ have totally different origins, the former being the signature of the Jahn-Teller effect while the latter is just a consequence of the ubiquitous octahedra rotations appearing in perovskites. RVO$_3$ are then unique materials in the sense that they are the only oxide perovskites with B=3d element showing a JT effect plus the sterically induced $Q_2^+$ mode, thereby they are the only member possessing entangled spin-orbital properties.

## IV. Conclusions

We have explained the modalities enabling a Jahn-Teller effect in ABX$_3$ perovskites and identified its signature: strongly localized electronic states are the key factor for a spontaneous electronic instability to produce a Jahn-Teller distortion with a specific octahedra deformation pattern that is experimentally detectable and detected. In materials with larger anions p and B cations d states hybridizations, there is no electronic instability that can break orbital degeneracies and the observed octahedra deformation has a distinct symmetry that is pushed by lattice mode couplings with rotations. The $Q_2^+$ octahedral deformation, appearing in LaMnO$_3$ or LaTiO$_3$, is not induced by an electronic instability and thus it is not a Jahn-Teller distortion. It is a consequence of octahedral rotations and tilts often appearing in the perovskite oxides due to pure semiclassical atomic size effects, i.e. geometric steric effects.

Our work provides a single theory explaining the Jahn-Teller effect and its specific signatures and reconcile the numerous experimental results of ABX$_3$ materials studied to date. Last but not least, our results settle the fact that dynamical correlations, and the Mott-Hubbard model, are absolutely not an essential aspect of the physics of ABX$_3$ materials showing insulating states despite the presence of degenerate states in the parent cubic cell, thereby defining DFT as a sufficient platform to study the physics of ABX$_3$ materials.

**Some details of the Method:** DFT calculations are performed with the VASP package [41,42] and electrons are nudged using the modified VASP routine[43]. We used PAW potentials with the outer 4s, 4p and 3d B cation electrons explicitly treated in the simulations. 6*6*4 k-point mesh was used for the relaxation of the (√2a, √2a, 2a) cubic cells, increased to 8*8*6 for plotting energy potential surfaces and seeking for OBS, accompanied by an energy cut-off of 500 eV. Energy potential surfaces were plotted with symmetry of the wavefunction off, in order to allow electrons to occupy the lowest energy state.

**Acknowledgments**: This work of JV has been supported by the European Research Council (ERC) Consolidator grant MINT under the Contract No. 615759. Calculations took advantages of the Occigen machines through the DARI project EPOC A0020910084 and of the DECI resource FIONN in Ireland at ICHEC through the PRACE project FiPSCO. The work of AZ was supported by Department of Energy, Office of Science, Basic Energy Science, MSE division under Grant No. DE-FG02-13ER46959 to CU Boulder. JV acknowledges support from A. Ralph at ICHEC supercomputers.